\documentclass[twocolumn,showpacs,preprintnumbers,amsmath,amssymb,superscriptaddress,floatfix]{revtex4}

\usepackage{graphicx}
\usepackage{dcolumn}
\usepackage{bm}

\begin{document}


\title{Metal to insulator transition in manganites - optical conductivity changes up to 24 eV}%

\author{A.~Rusydi}
\affiliation{Institut f\"{u}r Angewandte Physik,
Universit\"{a}t Hamburg, Jungiusstra$\ss$e 11, D-20355 Hamburg, Germany}

\author{R.~Rauer}
\affiliation{Dept. of Applied Physics, 
Chalmers University of Technology, 41296 G\"{o}teborg, Sweden}
 
\author{G.~Neuber}
\affiliation{Institut f\"{u}r Angewandte Physik,
Universit\"{a}t Hamburg, Jungiusstra$\ss$e 11, D-20355 Hamburg, Germany}

\author{M.~Bastjan}
\affiliation{Institut f\"{u}r Angewandte Physik,
Universit\"{a}t Hamburg, Jungiusstra$\ss$e 11, D-20355 Hamburg, Germany}

\author{I.~Mahns}
\affiliation{Institut f\"{u}r Angewandte Physik,
Universit\"{a}t Hamburg, Jungiusstra$\ss$e 11, D-20355 Hamburg, Germany}

\author{S.~M\"{u}ller}
\affiliation{Institut f\"{u}r Angewandte Physik,
Universit\"{a}t Hamburg, Jungiusstra$\ss$e 11, D-20355 Hamburg, Germany}

\author{P.~Saichu}
\affiliation{Institut f\"{u}r Angewandte Physik,
Universit\"{a}t Hamburg, Jungiusstra$\ss$e 11, D-20355 Hamburg, Germany}

\author{B.~Schulz}
\affiliation{Institut f\"{u}r Angewandte Physik,
Universit\"{a}t Hamburg, Jungiusstra$\ss$e 11, D-20355 Hamburg, Germany}

\author{G.~Stryganyuk}
\affiliation{II. Institut f\"{u}r Experimentalphysik,
Universit\"{a}t Hamburg, Luruper Chaussee 149, D-22761 Hamburg, Germany}

\author{K. D\"{o}rr}
\affiliation{Institut f\"{u}r Festk\"{o}rper- und Werkstofforschung,
D-01171 Dresden, Germany}

\author{G. A. Sawatzky}
\affiliation{Department of Physics and Astronomy, University of British Columbia, Vancouver, British Columbia V6T-1Z1, Canada}

\author{S.L. Cooper}
\affiliation{Department of Physics and Frederick Seitz Materials Research Laboratory, University of Illinois at Urbana-Champaign, USA}

\author{M.~R\"{u}bhausen}
\affiliation{Institut f\"{u}r Angewandte Physik,
Universit\"{a}t Hamburg, Jungiusstra$\ss$e 11, D-20355 Hamburg, Germany}

\date{\today}

\begin{abstract}

The electronic response of doped manganites at the transition from the paramagnetic insulating to the ferromagnetic metallic state in $\rm La_{1-x}Ca_{x}MnO_3$ for $\rm (x=0.3,0.2)$ was investigated by dc conductivity, ellipsometry, and VUV reflectance for energies between 0 and 24 eV. A stablized Kramers-Kronig transformation yields the optical conductivity and reveals changes in the optical spectral weight up to 24 eV at the metal to insulator transition. In the observed energy range, the spectral weight is conserved within $\rm 0.3 \%$. The redistribution of spectral weight between low and high energies has important ramifications for the down-folding of low-energy Hamiltonians. We discuss the importance of the charge-transfer, Coulomb onsite, Jahn-Teller, and screening effects to the electronic structure.
\end{abstract}


\maketitle

Among strongly correlated materials, the manganites exhibit a wealth of novel properties.  For example, some hexagonal insulating materials exhibit multiferroic behavior and the cubic doped manganites show charge ordering and the colossal magnetoresistance (CMR) effect.\cite{Millis96,Dagotto01} It is clear that the two key ingredients responsible for these diverse phenomena are, first, the high geometrical and spin frustration in the manganites, and second, the large number of competing interactions in these materials, the most important of which are the electron-electron and electron-phonon interactions.\cite{Millis96,Dagotto01,Allen1,Allen2,Kaliullin,Horsch,Khomski}  

There is disagreement as to which of these interactions is the primary driving force behind either the insulating phase of the manganites or the metal-to-insulator transition in the doped manganites.  Models of these phenomena commonly employ effective low-energy Hamiltonians, which predict new types of quasiparticle excitations, such as spin excitations, lattice polarons, spin polarons, or orbitons.\cite{Anderson,Millis96,Dagotto01,Allen2,Kaliullin,Horsch,Khomski,Liu,Yin06,Murakami,Saitoh,Kruger04,Gruninger} However, these effective Hamiltonians generally assume that the high energy degrees of freedom in these materials can be neglected, by 'down-folding' into a single band the large number of bands that characterize the actual system.  One example of this approach is the well-known Zhang-Rice singlet state of the cuprates.\cite{Rice88}  Another example is the Hamiltonian often employed to model a hole located between manganese and oxygen in the manganites:  a realistic model of this system would include both the Mn d-bands, which are split by the Coulomb interaction (Hubbard U), other interactions such as the Jahn-Teller effect, and  the oxygen 2p bands.  However, effective Hamiltonians used to describe this system typically ignore the oxygen p-bands, and consider only an effective d-band.\cite{Anderson,Millis96,Dagotto01,Allen2,Kaliullin,Horsch,Khomski,Goodenough}  Moreover, these effective low-energy models assume that there is no interaction between low and high energy degrees of freedom, which might be reflected, for example, in temperature- and/or doping-dependent spectral weight exchanges in the dielectric response between high and low energies. This might show that it is mandatory to consider the complex nature of the bandstructure explicitly.  Notably, experimental measurements of the dielectric response implicitly endorse the assumptions of effective low-energy models, as the dielectric response of these materials has, in the past, been typically measured only below 6 eV or lower, while the dielectric response above 6 eV is generally assumed to be insensitive to temperature and doping.\cite{Liu,Yin06,Saitoh,Kruger04,Gruninger,Okimoto97,Kovaleva04,Rauer06,Jung1,Jung2}

There is reason to question the assumption of effective low energy models, however.  If one considers the importance of local interactions in correlated materials, one might expect quite pronounced effects at higher energies due to the presence of locally unscreened interactions that lead to small-polaron, charge-transfer, or Mott-Hubbard physics.  In this case, one would expect local screening effects to induce strong changes at high energies through, for example, temperature- or doping-induced metal-to-insulator transitions.\cite{George,Phillips} 

In this study, we test the assumptions of not only effective low energy models of correlated oxides, but also of previous low energy dielectric measurements of these materials, by examining the temperature-dependent dielectric reponse of manganites over an unprecedented energy range, 0-24 eV. These measurements are made using a novel combination of dc-conductivity, spectroscopic ellipsometry, and VUV-reflectivity measurements, in order to examine the evolution of the optical conductivity from the paramagnetic insulating phase to the ferromagnetic metal phase of $\rm La_{0.7}Ca_{0.3}MnO_3$ and $\rm La_{0.8}Ca_{0.2}MnO_3$. Of particular interest to us are temperature-dependent spectral weight changes that occur in this energy range.  The integrated optical spectral weight is of special importance in studies of how a material's electronic structure evolves through a phase transition, as it is a conserved quantity.  Consequently, spectral weight changes through a phase transformation can reveal the relevant energy scales and the source of the free energy associated with that transformation. The main findings of this investigation are: (i) changes in the optical spectral weight exceed energies up to 24 eV in the metallic state - a significant fraction of the spectral weight of the Drude response stems from energies above 2 eV, presumably from electronic states centered at 8 and at 12 eV and (ii) in the spectral range between 0 and 24 eV the spectral weight is conserved within $\rm 0.3 \%$. Our results question the validity of effective low-energy one-band quasiparticle models, and demonstrate the importance of local interactions on the electronic structure of the manganites. We argue that screening effects of the Mott-Hubbard U which originate from the polarizability of the atoms and from the charge transfer between Mn and O are of crucial importance to understand the spectral weight changes over a wide energy range. The polarizabilities depend on the state of the system and therefore on temperature and doping.    

The experiments were performed on thin films, including (i) a 165 nm film of La$_{0.7}$Ca$_{0.3}$MnO$_3$ on orthorhombic $\rm NdGaO_3$ (110) and (ii) a 150 nm film of La$_{0.8}$Ca$_{0.2}$MnO$_3$ on orthorhombic $\rm NdGaO_3$ (110)  to minimize substrate related strain effects. The transition temperature $\rm T_c$ was 250 K ($\rm x=0.3$) and 175 K ($\rm x=0.2$), respectively. The detailed off-axis ablation process and the basic characterization of the films have been described elsewhere.\cite{Dorr} Four point resistance measurements were performed to measure the dc-conductivity. Spectroscopic ellipsometry measurements were performed in the spectral range between 0.5 eV and 6 eV by using the SE 850 (Sentech) ellipsometer and a UHV cryostat as described elsewhere.\cite{Neuber} Reflectance measurements in the high energy range between 4.5 and 24 eV we have used the Superlumi beamline at the DORIS storage ring of Hasylab (DESY) to measure the reflectance.\cite{Zimmerer} The calibration of the monochromator was done by measuring the luminescence yield of sodium salicylate ($\rm NaC_7H_5O_3$).\cite{Zimmerer} We outfitted the sample chamber with a gold mesh to measure the incident photon flux after the slit of the monochromator. The measurements at the superlumi beamline were very sensitive to small freeze out effects. We maintained a pressure of about $\rm 10^{-9}$ mbar, the temperature above 125 K, and we employed thermal cycling. 

\begin{figure}
\includegraphics[width=80.mm]{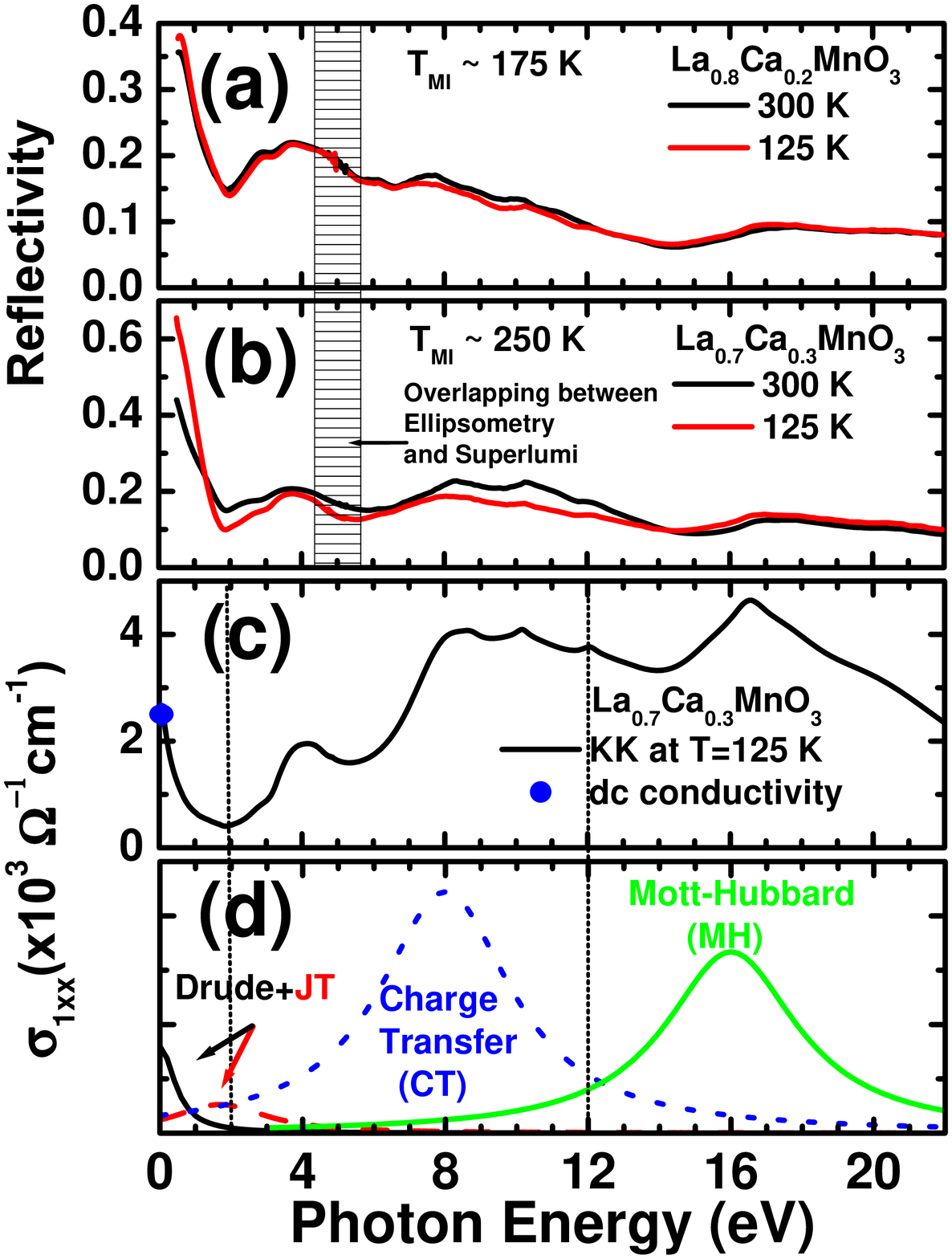}
\caption{Reflectivity from the self-normalizing ellipsometry setup and the superlumi beamline for temperatures above (black) and below (red) the metal to insulator transition for $\rm La_{0.8}Ca_{0.2}MnO_3$ (a) and $\rm La_{0.7}Ca_{0.3}MnO_3$ (b), respecively.  (c) optical conductivity at 125 K of $\rm La_{0.7}Ca_{0.3}MnO_3$ from a Kramers-Kronig transformation using the conductivity, relfectance, and ellipsometry results.(d) Schematic spectral regions: (i) Drude and Jahn-Teller (0-2 eV)(ii) the charge-transfer (2-12 eV), and (iv) local d-d transitions (12-22 eV). }
\end{figure}

Figure 1 shows the raw data as obtained by spectroscopic ellipsometry, VUV-reflectance, and conductivity measurements for La$_{0.7}$Ca$_{0.3}$MnO$_3$. Ellipsometry is a self-normalising technique, which makes it free from any ambiguities that are related to the normalisation of conventional reflectance results. Ellipsometry measures the real and imaginary part of the dielectric function. From this we can calculate the reflectance. We used ellipsometry-derived reflectance data between 4.5 and 5.5 eV to normalise the VUV reflectivity. In Fig. 1 (a) and (b), large changes in the reflectance at low and high energies are observed for $\rm La_{0.8}Ca_{0.2}MnO_3$ and $\rm La_{0.7}Ca_{0.3}MnO_3$ through the metal to insulator transition. The strongest changes occur at low energies as well as at rather high energies between 8 and 12 eV. The strength of the
changes increases with increased doping i.e., with the increased number of carriers. From these reflectivity measurements, it is quite clear that spectral weight redistribution associated with the metal-to-insulator transition is not adequately represented by reflectance measurements that are confined to low energies.  Further, it is clear from these data that one cannot assume that the high energy ($\rm >$ 2 eV) dielectric response is constant as a function of temperature, and indeed employing such an assumption in a simplified KK analysis will likely yield incorrect results for the redistribution of spectral weight across the phase transition\cite{Liu,Gruninger,Okimoto97,Jung1,Jung2}. We use for the Kramers Kronig transformation the dielectric function as obtained by our ellipsometry measurements in order to obtain the optical conductivity from the reflectance results shown in Fig. 1 (a) and (b). The dielectric function as obtained by ellipsometry contains a higher amount of information as compared to the reflectance. This procedure yields much more accurate results due to the application of both the large measured energy window and the use of the ellipsometry results. For the high energy cut-off, we have extrapolated our response by an $\rm \omega^{-4}$ power law. An example of the optical conductivity obtained here is shown in Fig. 1 (c) in the ferromagnetic metallic state of the manganites. In Fig 1 (c) we display the basic spectral response with peaks that are centered around 20 eV, 16 eV, 12 eV, 10 eV, 8 eV, 4 eV, 1 eV, and at the dc-limit in the metallic state. From Fig 1 (c) we identify three important spectral regions that we display in Fig 1(d). 

{\it Low-energy region - region I:} In the low energy region we identify a strong Drude response and an incoherent background up to about 2 eV. From this incoherent background the pseudogap develops in the insulating state of the manganites. This is the energy range where effective low-energy physics in terms of polarons or the orbital liquid dominates.\cite{Kaliullin,Horsch,Khomski} In the undoped material there is a clear evidence that the strong Jahn-Teller effect of the $\rm Mn^{3+}$ is important and results in an orthorhombic lattice distortion that couples for instance via an orbital exciton strongly to the electronic structure.\cite{Allen1,Allen2,Kruger04} The typical energy scale of these Jahn-Teller related excitations varies between 0.5 and 2 eV. In the low temperature limit the Jahn-Teller effect vanishes below $\rm T_c$.\cite{Bjornsson00} 

\begin{figure}
\includegraphics[width=80.mm]{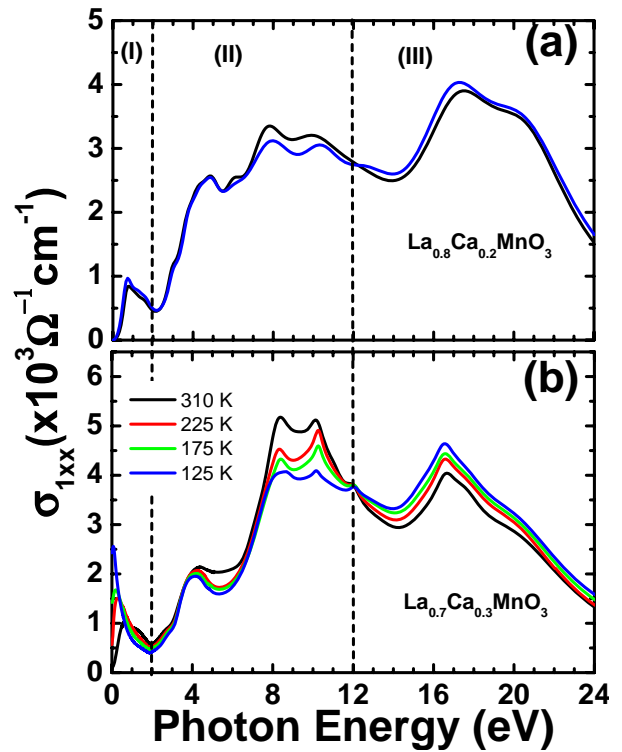}
\caption{
Optical conductivity in $\rm La_{0.8}Ca_{0.2}MnO_3$ and $\rm La_{0.7}Ca_{0.3}MnO_3$ as a function of temperature across the metal to insulator transition shown in (a) and (b), respectively. Note the spectral weight changes at high energies. }
\end{figure}

{\it The medium energy region - region II:} The charge-transfer excitations from the bands involving oxygen 2p states to the bands involving manganese d-states are found above 2 eV.\cite{Singh96} The local charge transfer
energy, to move an electron from the O(2-) ion to the Mn(3+) ion
is essentially given by the difference in the
ionization potential of O and the electron affinity of Mn in the actual solid. This involves the free ion values corrected for the very large Madelung potential reduced by the effects of covalency and the screening due to the polarizability of the soroundings. 
Studies of these effects estimate a charge transfer energy of 3 to 8 eV.\cite{Singh96,Eskes90} One can indeed observe that the transtion at around 5 eV gets pushed to lower energies as a function of doping and that the transitions at 8 and 12 eV eV get pushed towards higher energies as a function of doping. Furthermore, the crystal field splitting and the Jahn-Teller effect would split the energies of the charge-transfer excitations. The cubic crystal field splitting of about 2 eV results into two main charge-transfer bands of 4 and 8 eV. Due to the additional Jahn-Teller splitting of 0.5 to 1 eV we would expect four different charge-transfer bands: (i) O2p to Mn($\rm t_{2g}$  -JT); (ii) O2p to Mn($\rm t_{2g}$+JT); (iii) O2p to Mn($\rm e_{g}$-JT); (iV) O2p to Mn($\rm e_{g}$  +JT). Thus, we expect transitions at 3.5 eV, 4 eV, 8 eV, and 8.5 eV. 

{\it The high energy region - region III:} The highest energy scale in our problem is related to the bare energy of the Mn d to d transitions (unscreened Mott-Hubbard U). For electrons that are on one site, i.e. in the same orbital, one would expect such an energy to be of the order of 14.4 eV when calcuating the Coulomb energy $\rm e^2/(4 \pi \epsilon_0 r)$ with a maximum distance for the two charges of $\rm 10^{-10}$ m, i.e. the size of an Mn ion. This also implies that the on-site interaction is instantaneous on the sub femtosecond time scale. This is in qualitative agreement with Auger spectra on $\rm Cu_2O$ and cluster calcuations putting the $\rm U_{dd}$ to about 
10 eV.\cite{Ghijsen}  Considering the above mentioned energy scales, one would expect that the transtions are, split by the crystal field splitting, at 12 eV and 16 eV.  

Figure 2 shows the optical conductivity through the metal to insulator transition for (a) $\rm La_{0.8}Ca_{0.2}MnO_3$ ($\rm T_{MI} = 175 K$) and (b) $\rm La_{0.7}Ca_{0.3}MnO_3$ ($\rm T_{MI} = 250 K$). Changes in the optical conductivity increase as a function of doping. In particular, the strongly suppressed Drude response in the optical conductivity for $\rm La_{0.8}Ca_{0.2}MnO_3$ at the MI-transition is evident, leading to the most pronounced changes at higher energies around 8 to 12 eV. At 300 K, i.e., above the insulator to metal transition, the spectra show clear changes as a function of doping.  The screening terms would also change at the insulator to metal transition and as a function of doping as it can be seen when comparing Fig. 2 (a) and (b).

\begin{figure}
\includegraphics[width=80.mm]{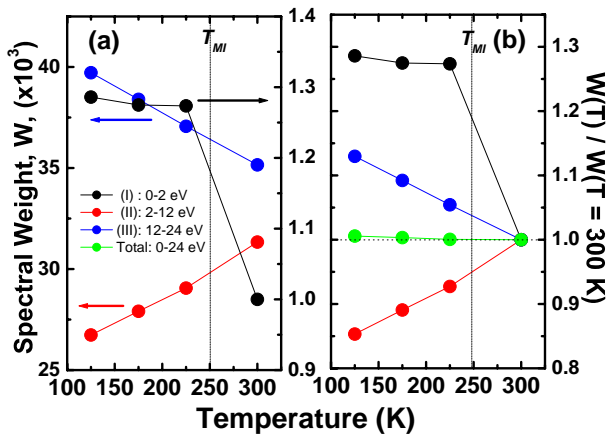}
\caption{
(a) integrated spectral weight of $\rm La_{0.7}Ca_{0.3}MnO_3$ in the three different spectral regions. (b) spectral weight changes in the different spectral regions and the overall integrated spectral weight. The spectral weight in the whole spectral region is conserved within $\rm 3/1000$, but that there is a spectral weight gain in the low energy region up to 2 eV.}
\end{figure}  

In the following we discuss the temperature dependence of the optical conductivity as it can be seen in Fig. 2 as well as in terms of the integrated spectral weight as it shown in Fig. 3. In Fig. 2 (a) and (b) we can find clearly strong renormalization effects as a function of temperature and shifts of spectral weight between the different fundamental processes as outlined in Fig. 1 (d). In order to discuss these effects more clearly we display the spectral weight changes of the three important energy regions in Fig. 3 (a) and (b): (i) low-energy region (0 - 2 eV), (ii) the medium energy region (2 to 12 eV), and (iii) seperated by an isosbestic point at 12 eV the high region (12 eV to 22 eV). It is important to note that the exchange of spectral weight between the Jahn-Teller and Drude peak is not sufficient to explain the strength of the spectral weight up to 2 eV in the metallic state, indicating that spectral weight gets transferred from much higher energies. We conclude that up to a third of the spectral weight comes from higher energies, i.e., presumably the charge transfer or Mott-Hubbard bands above 2 eV. The pronounced exchange of spectral weight at high energies between 10 eV and 24 eV could be related to the exchange of weight between spin dependent transitions in the ferromagnetic state. An optical transition between different Mn ions conserves the spin of the electron and is strongly dependent on the relative spin orientation of the ions. For ferro correlated spins the transition will involve high spin states in the final state of both ions, while for anit ferro correlated spins the transition will involve a low or intermediate spin state on one of the ions in the final state. The Hunds coupling would cost 3 times $\rm J_H$($\rm\approx 0.8 eV$) of about 2.4 eV when comparing these transitions, which would occur for d to d excitations.\cite{Kovaleva04}

In conclusion, we have measured the optical conductivity in an unprecedented energy range between 0 and 24 eV across the metal to insulator transition. By investigating spectral weight changes exceeding energies of more than 24 eV, we have identified three important spectral regions that we attribute to a Drude and Jahn-Teller, a charge-transfer response, and to Columb-d-d transitions. Our results put strong constraints on the interpretation of the physics in correlated materials such as the manganites. The effective low energy physics has to reflect appropiately the higher energy scales that are involved in the metal to insulator transition. Furthermore, we outline that experimental approaches limiting the temperature dependent studies of the reflectance to energies below 2 to 3 eV are questionable. 

We acknowlegde stimulating discussions with M.V. Klein, P. Abbamonte, P. Horsch, G. Kaliullin, and funding by DFG (RU773/3-1), HGF (VHZ-007), and BMBF.

\end{document}